\documentclass[mathematics,article,accept,pdftex,oneauthor,mathematics]{Definitions/mdpi} 
\firstpage{1} 
\pubvolume{1}
\issuenum{1}
\articlenumber{0}
\pubyear{2022}
\copyrightyear{2022}
\externaleditor{Academic Editor: 	Bo-Hao Chen}
\datereceived{16 September 2022} 
\dateaccepted{19 October 2022} 
\datepublished{} 
\hreflink{https://doi.org/} 

\Title{Residual Information in Deep Speaker Embedding Architectures}

\TitleCitation{Residual Information in Deep Speaker Embedding Architectures}

\Author{Adriana Stan
 \orcidA{}}

\AuthorNames{Adriana STAN}

\AuthorCitation{Stan, A.}

\address[1]{Department of Communications, Technical University of Cluj-Napoca,  400114, Cluj-Napoca, Romania; adriana.stan@com.utcluj.ro
}

\abstract{Speaker embeddings represent a means to extract representative vectorial representations from a speech signal such that the representation pertains to the speaker identity alone. The embeddings are commonly used to classify and discriminate between different speakers. However, there is no objective measure to evaluate the ability of a speaker embedding to disentangle the speaker identity from the other speech characteristics. This means that the embeddings are far from ideal,  highly dependent on the training corpus and still include a degree of residual information pertaining to factors such as linguistic content, recording conditions or speaking style of the utterance. 
This paper introduces an analysis over six sets of speaker embeddings extracted with some of the most recent and high-performing deep neural network (DNN) architectures, and in particular, the degree to which they are able to truly disentangle the speaker identity from the speech signal. To correctly evaluate the architectures, a large multi-speaker parallel speech dataset is used. The dataset includes 46 speakers uttering the same set of prompts,  recorded in either a professional studio or their home environments.
The analysis looks into the intra- and inter-speaker similarity measures computed over the different embedding sets, as well as if simple classification and regression methods are able to extract several residual information factors from the speaker embeddings.  
The results show that the discriminative power of the analyzed embeddings is very high, yet 
across all the analyzed architectures, residual information is still present in the representations in the form of a high correlation to the  recording conditions, linguistic contents and utterance duration. However, we show that this correlation, although not ideal, could still be useful in downstream tasks.  
The low-dimensional projections of the speaker embeddings show similar behavior patterns across the embedding sets with respect to intra-speaker data clustering and utterance outlier detection. }

\keyword{speaker embeddings; x-vectors; deep representations; deep embeddings; speaker disentanglement; speaker recognition; residual information; neural architectures; deep learning; artificial intelligence} 

\MSC{68T01; 68T07
}

\begin{document}

\section{Introduction}

Recorded speech is an inherently complex signal including information related to the linguistic contents, prosodic or style factors (such as rhythm and intonation), as well as speaker characteristics (such as physiological traits, gender, ethnicity or social background). Humans have the ability to disentangle almost all of these factors and extract their abstractions, being able to reproduce and recognize similar patterns across spoken data from different sources. An essential part of the speech signal with a multitude of downstream applications is related to the disentanglement of the speaker identity. Such an accurate representation of the speaker identity would make it extremely useful in tasks such as speaker recognition and verification applications, text-to-speech synthesis and voice cloning~\cite{transferlearning_tts}, anonymization or generating new, unseen speaker identities~\cite{spk_gen}.  
There is already a large number of published works which focus on speaker discrimination, meaning that their task is to estimate if two or more acoustic signals pertain to the same speaker identity. However, in terms of accurately representing the speaker for generative processes (such as text-to-speech or voice cloning), to  date, there are no published methods which can accurately solely represent the speaker identity and disregard other factors related to the acoustic signal, such as recording conditions or linguistic contents, although many of these applications use the derived representations as input or conditioning. It is therefore essential to perform an analysis of how well the current speaker representations model the speaker identity. 

As a result, in this paper, the focus is on finding out how much \textit{residual information} (i.e., not pertaining to the speaker identity) is present in various deep speaker embeddings. Six open source, easy-to-use, readily available implementations were selected. The implementations report the state-of-the-art results for speaker classification and diarization tasks. A first evaluation carried out in this work aimed at directly comparing the architectures' performances with respect to their intended use, i.e., speaker discrimination. The equal error rates (EER) and inter- and intra-speaker similarity measures were computed over a large multi-speaker parallel dataset. In the second step of the evaluation, the architectures' derived speaker embeddings were analyzed in terms of the amount of residual information present within them, such as the utterance duration, signal-to-noise ratio, linguistic contents and recording conditions. Simple classification and regression algorithms were employed in an attempt to extract this information from the representations. The results show that, to a large extent, the embeddings exhibit a high dependency on these factors, and as such, the speaker identity is not truly disentangled. 
Based on these initial results, we then attempted to explore if this residual information could still be useful in downstream tasks. The derived speaker embeddings were plotted in a low-dimensional representation to verify if  they exhibit similar patterns with respect to clustering different recording sessions and background conditions, as well as to separate utterance outliers and ill-behaved speakers.  
Such information could be exploited for selecting the appropriate speakers and a set of samples for a text-to-speech synthesis system or data augmentation process in multi-speaker systems. 

The \textbf{contributions} of this paper can be summarized as follows:
\begin{itemize}
\item{Six of the most recent speaker embedding deep neural networks are directly compared with respect to their discriminative and generative characteristics;}
\item{The analysis is carried out over a parallel dataset consisting of 46 speakers uttering the same prompts;}
\item{The equal error rates (EER) and inter- and intra-speaker similarity measures for the six architectures are evaluated;}
\item{Decision trees and light gradient boosting machine algorithms are employed to evaluate the amount of residual information present in the embeddings;}
\item{Low-dimensional tSNE-based representations of the embedding space for the six architectures are evaluated in terms of outlier detection and intra-speaker data clustering.}
\end{itemize}

The paper is organized as follows: Section \ref{sec:rel-work} presents some of the previous studies which address the development of accurate speaker embeddings, as well as their use in voice cloning and text-to-speech synthesis systems. Section \ref{sec:exper} describes the audio data and speaker embedding architectures adopted in this work. The results of the evaluation are shown in Section~\ref{sec:results}, while the conclusion and discussions are introduced in Section~\ref{sec:conc}.

\section{Related Work}
\label{sec:rel-work}



Speaker recognition has been the focus of the research community for quite a long time now, as it is essential for identification, verification and diarization tasks. Speaker identification refers to determining the identity of a spoken utterance from a set of predefined speakers. Speaker verification aims to predict if two utterances pertain to the same speaker or not, while speaker diarization is targeted at separating the different speaker identities present in a larger audio clip and assigning each audio segment to the corresponding speaker. Although these three sub-tasks of speaker recognition seem  different in principle, they all share the common component of extracting the numeric representations able to accurately depict individual speaker identities.

Some of the first methods for speaker recognition were based on spectral and template matching, commonly using Mel-Frequency Cepstral Coefficients~\cite{Luck1969AutomaticSV, 1163530}, Linear Prediction Coefficients (LPC)~\cite{wenxing,chandra} or Perceptual Linear Prediction (PLP) Coefficients~\cite{mclaren,plchot}. 
Starting with the 1990s, Gaussian Mixture Models (GMM)~\cite{gmm,365379} became more prevalent. The models estimated the statistics of various speech signal representations for each individual speaker within the dataset. The verification or recognition was performed by computing the distance between the target speaker and each of the probability distribution functions within the GMMs set. With the addition of the Universal Background Model (UBM)~\cite{REYNOLDS200019} and Support Vector Machines (SVM)~\cite{1618704}, the performances of speaker recognition methods kept improving. Yet, the evaluations were performed on small, curated, clean datasets and more than often in text-dependent scenarios.
Dimensionality reduction techniques were subsequently applied so as to extract the axes of maximum variation among the speakers of interest. Within this area, Principal Component Analysis (PCA)~\cite{1223488} and i-vectors~\cite{5545402} rapidly gained popularity. 

The major improvements in speaker recognition, as in many other application fields, came from the introduction of deep learning architectures able to abstractize the information present in the speech signal benefiting from a large amount of spoken data. The first step toward the DNN-based representations was simply to use the deep architecture's posterior probabilities instead of the GMM-based ones~\cite{6026951}. Similar to the PCA technique in traditional speaker recognition models, DNN-based bottleneck features became popular~\cite{Hinton2006ReducingTD}, being called d-vectors and extracted at the frame level. D-vectors are part of a larger category of DNN-based representations, called embeddings. 
X-vectors are also embeddings extracted with Time-Delay Neural Networks (TDNN)~\cite{8461375, GarciaRomero2019xVectorDR, You2019MultiTaskLW} at the segment level, and they became the standard method for speaker recognition applications. Some other deep architectures employed in speaker recognition are RawNet~\cite{rawnet,rawnet2} and ResNet~\cite{chung2020in,kwon2021ins,thienpondt2020idlab}. These types of embeddings are extracted in an end-to-end manner, meaning that the network is in charge of both finding adequate representations as well as determining the final decision related to the speaker-related task. Previous methods used either the Probabilistic Linear Discriminant Analysis (PLDA) or cosine similarity to estimate the similarities or dissimilarities between the output representations.
Some recent studies even attempted to adapt other speech-based neural representations for speaker recognition~\cite{9746952}. An extended overview of the deep learning-based speaker embedding representations can be found in \cite{BAI202165,9458301}.

As more and more methods were published, a common evaluation benchmark was required to correctly compare their individual results. Several speaker recognition workshops and challenges have been organized, such as the NIST Speaker Recognition 
Evaluation.
 ({\url{https://www.nist.gov/itl/iad/mig/speaker-recognition}, accessed on 22 October 2022}), Odyssey 
 ({\url{http://www.odyssey2022.org/}, accessed on 22 October 2022}) or VoxSRC.
 ({\url{https://www.robots.ox.ac.uk/~vgg/data/voxceleb/}\linebreak \url{interspeech2022.html}, accessed on 22 October 2022}). Within the 2022 VoxSRC challenge, there were two main tracks related to speaker verification (open and closed sets) and speaker diarization. The best performing systems included ResNet and ECAPA-TDNN architectures augmented with self-supervised learned (SSL) representations of the audio signal~\cite{ravana, cai2022kriston, suh2022returnzero,hccl}.

Although the methods described above are aimed solely at speaker recognition, verification and diarization applications, their findings can be applied to other speech-related tasks. One of the most important and widely used is that of speech synthesis and voice cloning. Speaker embeddings extracted from networks trained on a large number of speakers can be used to condition multi-speaker synthesis models. Using externally learned embeddings enables the models to perform a fast or zero-shot adaptation for unseen speakers~\cite{zs,9054535, 9606610, 9383454, 9380685}. And it is for these tasks that a more elaborate analysis of the embeddings' accuracy is extremely important.

\section{Experimental Setup}
\label{sec:exper}

\subsection{Speech Data}
\label{sec:audio}

A problematic part of the speaker embeddings' evaluation is the fact that the spoken data across the speakers may vary. This means that there is a possibility that the number and linguistic content of each speaker's utterance subset may influence the results. Therefore, in this study, we used one of the largest parallel spoken datasets available. The dataset is the extended version of the SWARA corpus~\cite{stan_sped2017}. The initial version of the corpus---which will be referred to as SWARA1.0
---includes 18 speakers recorded in a professional studio. Each speaker read aloud between 921 and 1493 utterances. This dataset was recently extended with an additional 28 speakers---we will refer to this subset as SWARA2.0. However, due to the COVID-19 pandemic, the recordings were performed in the speakers' home environments with semi-professional equipment. The speakers read between 1597 and \mbox{1797 utterances}. As the SWARA2.0 was recorded in home conditions, we expect the background noise and reverberation to affect the performance of the embeddings extracted from this dataset. 

In both SWARA1.0 and SWARA2.0 subsets, the speakers were provided with the same text prompts to be read aloud. However, due to the lack of control, especially in the SWARA2.0 scenario, only 712 utterances are truly parallel across all 46 speakers. This means that for the rest of the utterances, the speakers either made deletions, insertions or substitutions with respect to the prompt or did not record some of the prompts at all. 
There are 24 female speakers and 22 male speakers in the combined datasets, which amount to 32.752 utterances with a total duration of 38 h and 29 min. All data were resampled at 16~kHz and start and end silence segments were trimmed.

\subsection{Speaker Embedding Networks}

Numerous studies focused on extracting deep learning-based representations for speaker characteristics. Most of these studies are, of course, aimed at discriminating between speakers and performing accurate recognition and diarization tasks. 
For our evaluation, we targeted the DNN-based architectures which are open source, easily accessible and usable and also provide good pre-trained models. The following architectures were selected:


(1) \textbf{Pyannote} ({\url{https://github.com/pyannote/pyannote-audio}, accessed on 22 October 2022}) is an open-source toolkit written in Python for speaker diarization~\cite{Bredin2020, Bredin2021}. It uses a SincNet~\cite{sincnet} architecture, followed by a series of TDNN layers and an average pooling layer. It also includes the implementations for Speech Brain and NeMo Titanet architectures, but we did not use them in this study.


(2) \textbf{Speech Brain's}~\cite{speechbrain} speaker verification network.
 ({\url{https://github.com/speechbrain/speechbrain}, accessed on 22 October 2022}) is based on the ECAPA-TDNN~\cite{ecapatdnn} model. It includes a sequence of convolutional and residual blocks, using the additive margin softmax loss as training objective. The speaker embeddings are formed using attentive statistical pooling.

(3) \textbf{Clova AI} \cite{chung2020in,kwon2021ins} uses a ResNet-like architecture.
, ({\url{https://github.com/clovaai/voxceleb_trainer}, accessed on 22 October 2022}) and similarly averages the frame-level representations in the final embeddings. The difference is in the change in objective function and the use of training data augmentation.

(4) \textbf{NeMo Titanet} ({repository for all NeMo architectures is available here:.
 \url{https://github.com/NVIDIA/NeMo/}, accessed on 22 October 2022}) uses the Titanet architecture~\cite{titanet} which is based on ContextNet~\cite{contextnet}. The model uses 1D depth-wise separable convolutions with squeeze-and-excitation layers. The output embeddings are obtained by averaging the statistics of the intermediate variable-length representations.

(5) \textbf{NeMo SpeakerNet} uses an ASR architecture's encoder, namely QuartzNet~\cite{Koluguri2020SpeakerNet1D}, as a high-level feature extractor and averages these features within a pooling layer so as to capture the time-independent speaker features. 

(6) \textbf{NeMo ECAPA-TDNN} is similar to the Speech Brain architecture and uses the ECAPA-TDNN structure~\cite{Dawalatabad2021ECAPATDNNEF}. The difference is that, instead of the residual blocks, the NeMo implementation uses group convolution blocks of single dilation.

Because we aim to use the embeddings in speech synthesis systems, no fine-tuning was performed over the available pre-trained models. Fine-tuning would imply that any time a new dataset is used for synthesis, the embedding networks would need to be re-trained, which would not be feasible. Therefore, we explore the architecture's behavior as is and as their authors made them available for the wide research community.

\section{Evaluation}
\label{sec:results}

We base our evaluation on a set of analyses which we consider relevant for the use of the speaker embeddings in downstream tasks, outside the speaker recognition or discrimination applications. The following subsections introduce the results of the evaluation scenarios. 

\subsection{EER, Intra- and Inter-Speaker Similarity}

Being derived from neural architectures aimed at speaker recognition, thus in a discriminative-oriented task, the speaker embeddings' performance is commonly evaluated in terms of the equal error rate (EER). The EER is defined as the point on the ROC curve where the false acceptance rate (FAR) is equal to the false rejection rate (FRR). The threshold to compute the EER is generally based on the cosine similarity between pairs of speaker embeddings, defined as:

\begin{equation}
\cos ({\bf e}_1,{\bf e}_2)= {{\bf e}_1 {\bf e}_2 \over \|{\bf e}_1\| \|{\bf e}_2\|} = \frac{ \sum_{i=1}^{n}{{\bf e}_{1,i}{\bf e}_{2,i}} }{ \sqrt{\sum_{i=1}^{n}{({\bf e}_{1,i})^2}} \sqrt{\sum_{i=1}^{n}{({\bf e}_{2,i})^2}} }
\end{equation}

Because we are using a new speech dataset, it is essential to first evaluate the targeted performance of the selected architectures.
Therefore, the first step of our analysis involved the computation of the EER results across the architectures, speakers and sets of speakers. 
A set of 46,000
 random pairs of utterances were selected. The random pairs were chosen such that each speaker is present in 1000 pairs, of which 200 are same-speaker pairs. The EER results are shown in Table~\ref{tbl:eer}. We evaluate the results over the entire set of 46,000 pairs and separately for pairs containing only samples for the female speakers, male speakers, speakers from the SWARA1.0 subset and speakers from the SWARA2.0 subset. It can be noticed that all architectures achieve an EER below 1\% and that female speakers are better discriminated than male speakers. The same is true for the SWARA1.0 speakers versus the SWARA2.0 speakers. The best results are obtained by the NeMo Titanet architecture. 

\begin{table}[H] 
\caption{EER values for the different speaker embedding architectures and speaker subsets. Arrows mark the direction of best performance. Best results are highlighted in boldface. \label{tbl:eer}}
\newcolumntype{C}{>{\centering\arraybackslash}X}
\begin{tabularx}{\textwidth}{lCCCCC}
\toprule
\textbf{Architecture}	& \textbf{All\boldmath{$\downarrow$}} & \textbf{Female\boldmath{$\downarrow$}} & \textbf{Male\boldmath{$\downarrow$}} & \textbf{SWARA1.0\boldmath{$\downarrow$}} & \textbf{SWARA2.0\boldmath{$\downarrow$}}\\
\midrule
Pyannote		    & 0.040	         & 0.055 & 0.039 & 0.024 & 0.047 \\
Speech Brain		& 0.025          & 0.027 & 0.031 & 0.011 & 0.031\\
Clova AI            & 0.055          & 0.060  & 0.081      & 0.031      & 0.073      \\
NeMo Titanet		& \textbf{0.018} & \textbf{0.014} & \textbf{0.027} & \textbf{0.005} & \textbf{0.024} \\
NeMo SpeakerNet		& 0.039  & 0.045 & 0.051 & 0.024 & 0.048\\
NeMo ECAPA-TDNN		& 0.032  & 0.035 & 0.041 & 0.023 & 0.038\\
\bottomrule
\end{tabularx}
\end{table}

\begin{figure}[H]
\includegraphics[width=\columnwidth]{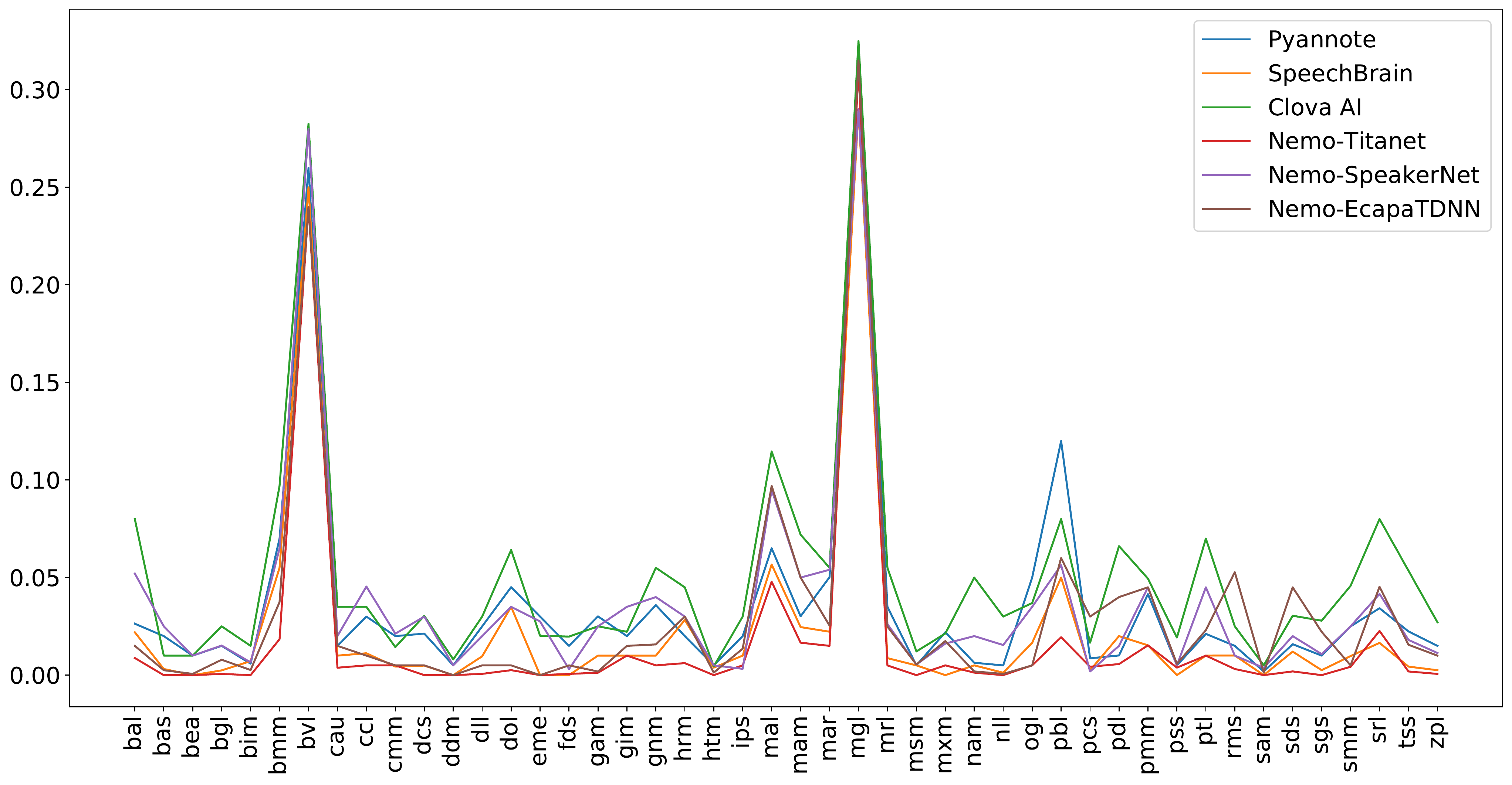}
\caption{EER values for the individual speakers across the embedding architectures.\label{fig:spk-eer}}
\end{figure}

NeMo Titanet is also the best performing architecture for each individual speaker (see Figure~\ref{fig:spk-eer}). 
However, speakers \texttt{bvl}, \texttt{mgl} and \texttt{pbl} have significantly higher EER values than the rest of the speakers. When analyzing these speakers' recordings, we noticed that indeed the background conditions are considerably poorer than for the other speakers and that, in some cases, the segmentation of the waveform is performed after or before the end of the utterance. 

\begin{figure}[H]
\includegraphics[width=\columnwidth]{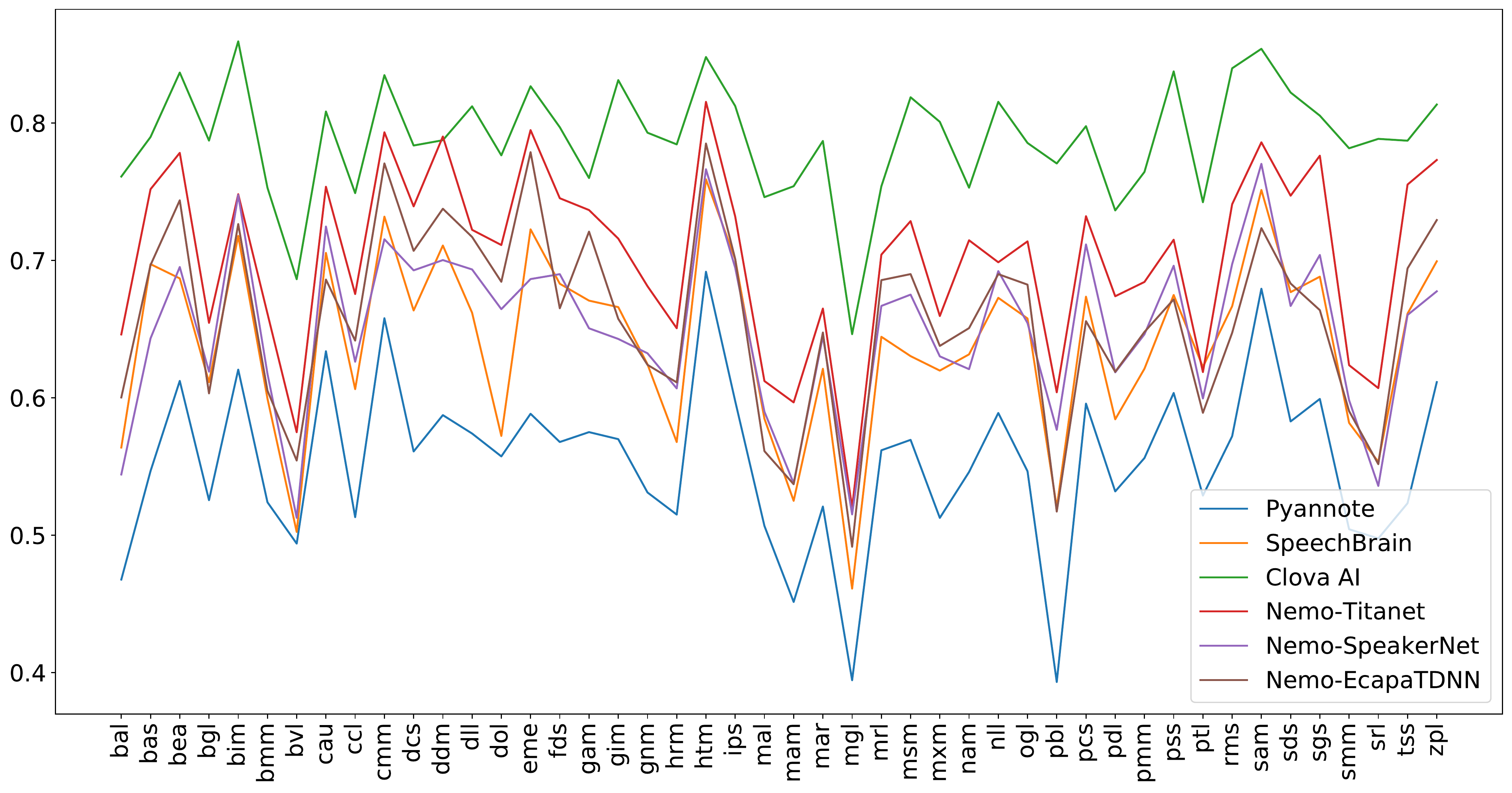}
\caption{Intra-speaker cosine similarity for each speaker and embedding architecture.\label{fig:intra-speaker-sim}}
\end{figure}

The EER gives the optimum threshold for which the $FAR$ is equal to the $FRR$, but it does not detail the accuracy of the representation. For downstream tasks, the inter- and intra-speaker similarity measures would be more informative. This means that the discrimination between speakers can be translated into high intra-speaker similarity and low inter-speaker similarity, while any intermediate values should ideally represent perceptually similar speakers. The intra-speaker similarity values are presented in Figure~\ref{fig:intra-speaker-sim}. The similarity was computed over all pairs of utterances from the same speaker averaged by their number. It can be noticed that the Clova AI architecture exhibits the highest intra-speaker similarity measure, and that again, speakers \texttt{bvl, mgl} and \texttt{pbl} have the lowest intra-speaker similarity. The least performing architecture is Pyannote. 
In Table~\ref{tbl:intra-spk}, we average the above scores at the system-level and also across the different speaker subsets. The Clova AI, Female and SWARA1.0 speakers exhibit the highest intra-speaker similarity measures.

\begin{table}[H] 
\caption{Average \textbf{intra-speaker} cosine similarity across the different speaker embedding architectures and different speaker subsets. Arrows mark the direction of best performance. Best results are highlighted in boldface.\label{tbl:intra-spk}}
\newcolumntype{C}{>{\centering\arraybackslash}X}
\begin{tabularx}{\textwidth}{lCCCCC}
\toprule
\textbf{Architecture}	& \textbf{All\boldmath{$\uparrow$}} & \textbf{Female\boldmath{$\uparrow$}}  & \textbf{Male\boldmath{$\uparrow$}} & \textbf{Swara1.0\boldmath{$\uparrow$}} & \textbf{Swara2.0\boldmath{$\uparrow$}} \\
\midrule
Pyannote		  &  0.554 & 0.557 & 0.550 & 0.589 & 0.531\\
Speech Brain	  &	0.640 & 0.651 & 0.629 & 0.686 & 0.610\\
Clova AI          & \textbf{0.788} & \textbf{0.790} & \textbf{0.786} & \textbf{0.810} & \textbf{0.774} \\
NeMo Titanet	  &	0.702 & 0.711 &  0.695& 0.750 & 0.672\\
NeMo SpeakerNet	  &	0.651 & 0.658 & 0.644 & 0.693 & 0.623\\
NeMo ECAPA-TDNN	  &	0.658 & 0.670 & 0.647 & 0.696 & 0.635\\
\bottomrule
\end{tabularx}
\end{table}

\begin{figure}[H]
\includegraphics[width=0.8\columnwidth]{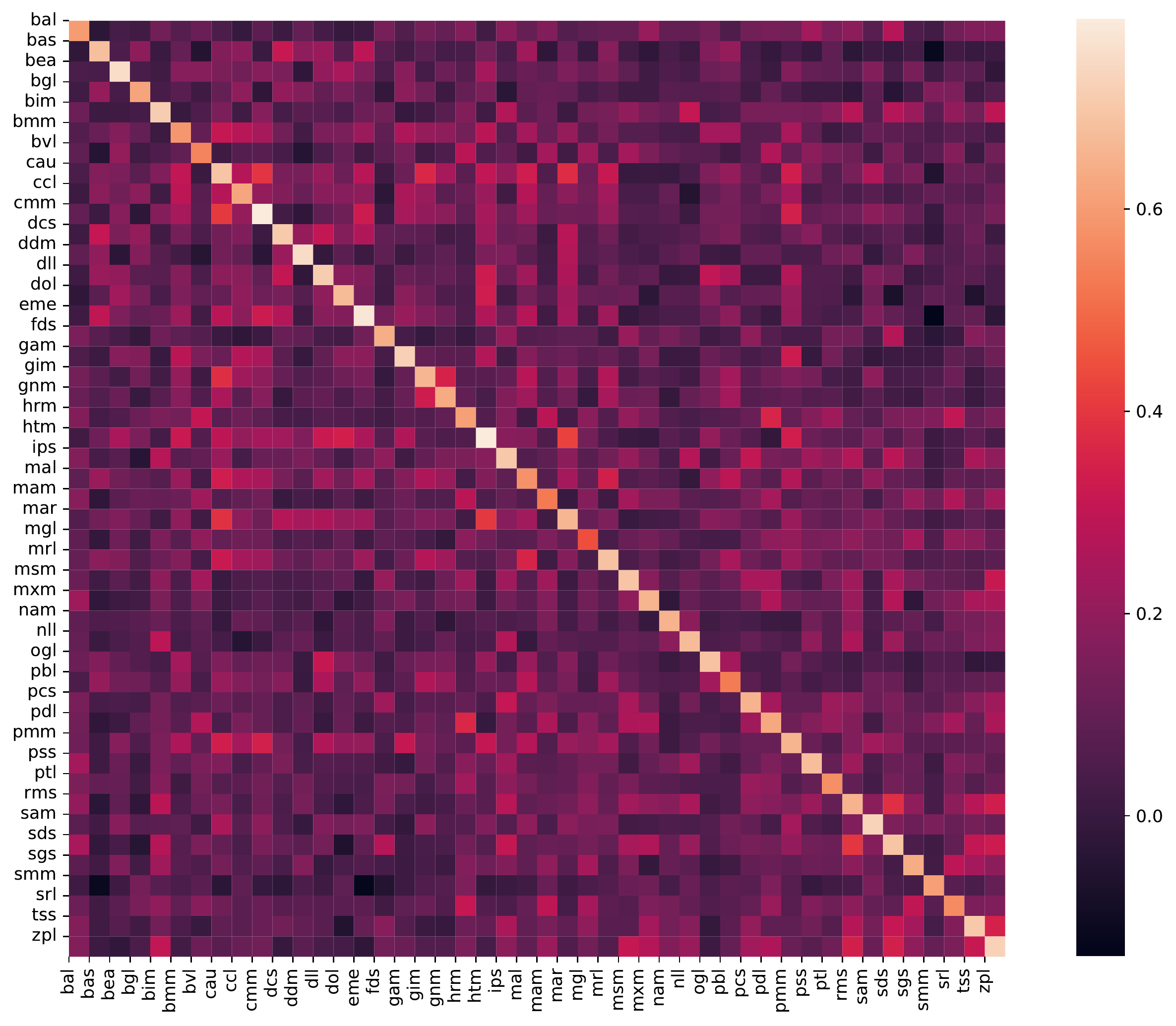}
\caption{Inter-speaker similarity matrix for the NeMo ECAPA-TDNN architecture.\label{fig:inter-speaker-sim}}
\end{figure}

For the inter-speaker similarity, we use the 46,000 random pairs of utterances used to evaluate the EER and select only those which pertain to different speaker identities. We then compute the average cosine similarities between each pair of speakers. In this scenario,
we would expect the architectures to exhibit very low values so as to maximize the discriminative characteristics of the representation. In Table~\ref{tbl:inter-speaker}, we introduce these results averaged across the architectures and speaker subsets. 
Although NeMo Titanet seemed to show the best performance in the previous tasks, in terms of discriminative power, NeMo ECAPA-TDNN is the most efficient (with an exception for the inter-male speakers where Pyannote is best). We show the inter-speaker similarity matrix for the NeMo ECAPA-TDNN architecture in Figure~\ref{fig:inter-speaker-sim}. The closest speakers based on these scores are \texttt{htm} and \texttt{mar}, with a similarity of 0.42, followed by \texttt{cmm} and \texttt{cau} at a 0.40 similarity. These pairs are female speakers and do indeed have perceptually similar voices. 

In this section, we looked at common measures to evaluate speaker embedding architectures while aiming to extract additional information that may be useful for downstream tasks. For example, inter-speaker similarity could be used to train speech synthesis systems in limited data scenarios, where data augmentation can be performed by using speech from a different, yet similar sounding speaker.

\begin{table}[H] 
\caption{Average \textbf{inter-speaker
} cosine similarity across the different speaker embedding architectures and different speaker subsets. Arrows mark the direction of best performance. Best results are highlighted in boldface.\label{tbl:inter-speaker}}
\newcolumntype{C}{>{\centering\arraybackslash}X}
\begin{tabularx}{\textwidth}{lCCCCC}
\toprule
\textbf{Architectures}	& \textbf{All\boldmath{$\downarrow$}} & \textbf{Female\boldmath{$\downarrow$}}  & \textbf{Male\boldmath{$\downarrow$}} & \textbf{Swara1.0\boldmath{$\downarrow$}} & \textbf{Swara2.0\boldmath{$\downarrow$}} \\
\midrule
Pyannote		  & 0.127 & 0.195 & \textbf{0.129} & 0.139 & 0.127\\
Speech Brain	  &	0.122 & 0.188 & 0.143 & 0.139 & 0.118\\
Clova AI          & 0.133 & 0.188 & 0.157  & 0.142  & 0.135 \\ 
NeMo Titanet	  &	0.132 & 0.187 & 0.156 & 0.143 & 0.135\\
NeMo SpeakerNet	  &	0.195 & 0.272 & 0.226 & 0.199 & 0.198\\
NeMo ECAPA-TDNN	  &	\textbf{0.107} & \textbf{0.151} & 0.140 & \textbf{0.136} & \textbf{0.110}\\
\bottomrule
\end{tabularx}
\end{table}

\unskip

\subsection{Speaker Identity Disentanglement}

The previous set of results are definitely aimed at providing the best representation for speaker classification tasks. However, in many downstream applications of the speaker embeddings, this is not enough. And this is true especially in speech synthesis systems, where the embeddings are used as additional input, while information related to the linguistic content and prosodic patterns are the main inputs. 
Therefore, if information pertaining to other aspects of the speech is present in the embeddings, this can lead to unwanted effects and bias within the training procedure. 
Starting from the above statement, in this section, we want to explore how much residual information is present within the speaker embeddings, and thus if the speaker identity is truly disentangled from all other speech factors. 

A simple method to determine the presence of residual information is to see if simple machine learning algorithms are able to extract this information from the embeddings. 
We adopt two separate algorithms, depending on the task: a decision tree for the classification tasks and a LightGBM~\cite{lightgbm} for the regression tasks. The two algorithms were chosen as they are some of the simplest, yet most powerful traditional machine learning methods, and their results are comparable across tasks and datasets. The same parameters were used across the tasks, and the speech dataset was randomly split into 80\% training and 20\% test sets.

Table~\ref{tbl:f1scores} shows the results expressed in F1-scores for the classification of speaker gender and speaker identity. For these two targets, the results are supposed to be high, as the embeddings should incorporate this information. 
Two other targets shown in the table are the text length---expressed in number of characters---and the recording conditions. The recording condition is encoded as a binary classification for the two subsets, SWARA1.0 and SWARA2.0. For these two columns, the accuracy of the predictions should be limited, as information about these two targets should not be easily extracted from the embeddings. For the text length, the results are as expected, and the number of characters present in the utterance cannot be extracted from the embeddings. Yet,
an interesting result occurs in the recording conditions column, where the prediction accuracy is rather high. This means that the additional spectral artefacts present in the home recordings are indeed influential for the final embeddings across all embedding architectures.

\begin{table}[H] 
\caption{F1-scores for various classification tasks. Arrows mark the direction of best performance. Best results are highlighted in boldface. \label{tbl:f1scores}}
\newcolumntype{C}{>{\centering\arraybackslash}X}
\begin{tabularx}{\textwidth}{lCCCC}
\toprule
\textbf{Architecture}	& \textbf{Speaker ID\boldmath{$\uparrow$}}& \textbf{Gender\boldmath{$\uparrow$}} &\textbf{Utterance Duration\boldmath{$\downarrow$}} & \textbf{Recording Condition\boldmath{$\downarrow$}} \\
\midrule
Pyannote		    & 0.76	& 0.94 & 0.016	& \textbf{0.87} \\
Speech Brain		& 0.84	& 0.96 & 0.011	& 0.90 \\
Clova AI           & 0.85  & 0.98 & 0.015  & 0.92 \\
NeMo Titanet		& \textbf{0.90}	& \textbf{0.97} & \textbf{0.010}  & 0.95 \\
NeMo SpeakerNet		& 0.85	& 0.96 & 0.014  & 0.91 \\
NeMo ECAPA-TDNN		& 0.87	& 0.96 & 0.015	& 0.95 \\
\bottomrule
\end{tabularx}
\end{table}

For the regression tasks, we look at the utterance duration (measured in seconds), the signal-to-noise ratio (SNR) and the linguistic contents. The SNR was computed with the WADA~\cite{wada} algorithm. For the text contents, we are only looking at the utterance id, assuming that similar characteristics would be present across speaker embeddings for the same linguistic contents. The accuracy of the LightGBM is measured in terms of the Spearman Rank Correlation Coefficient (SRCC) of the predicted values versus the target values. Table~\ref{tbl:srccscores} shows the results, and they are not encouraging. All the architectures show a high correlation factor (>0.7) with respect to the evaluated targets. This means that this type of information is not efficiently disentangled from the resulting speaker embedding. 

A separate regression task looked into the prediction of the average F0 value at the utterance level (the last column in Table~\ref{tbl:srccscores}). The correlations are rather high and the average mean squared error for the six architectures is 12 Hz. 

\begin{table}[H] 
\caption{LightGBM-based SRCC results for various regression tasks. Arrows mark the direction of best performance. Best results are highlighted in boldface.\label{tbl:srccscores}}
\newcolumntype{C}{>{\centering\arraybackslash}X}
\begin{tabularx}{\textwidth}{lCCCC}
\toprule
\textbf{Architecture}	& \textbf{Utterance Duration}\boldmath{$\downarrow$}  & \textbf{SNR}\boldmath{$\downarrow$} & \textbf{Linguistic Contents}\boldmath{$\downarrow$}  & \textbf{F0}\boldmath{$\uparrow$}\\
\midrule
Pyannote		    & 0.830 & 0.771 & 0.723 & 0.958\\
Speech Brain		& 0.734 & 0.749 & 0.742 & 0.959\\
Clova AI            & 0.750 & 0.791 &  0.734 & \textbf{0.976} \\
NeMo Titanet		& \textbf{0.704} & \textbf{0.747} & 0.798 & 0.964\\
NeMo SpeakerNet		& 0.796 & 0.758 & \textbf{0.709} & 0.958\\
NeMo ECAPA-TDNN		& 0.862 & 0.787 & 0.775 & 0.962\\
\bottomrule
\end{tabularx}
\end{table}

The results in this section showed that although the task of the embedding architectures is to represent the speaker identity as accurately as possible, residual information is still present within them and that more work should be performed to find a more suitable representation for the downstream tasks. In the following section, we would like to explore how we can use this residual information to the benefit of other tasks by using visual representations of the embeddings.

\subsection{Visual Representations}

\begin{figure}[H]
\includegraphics[width=\columnwidth]{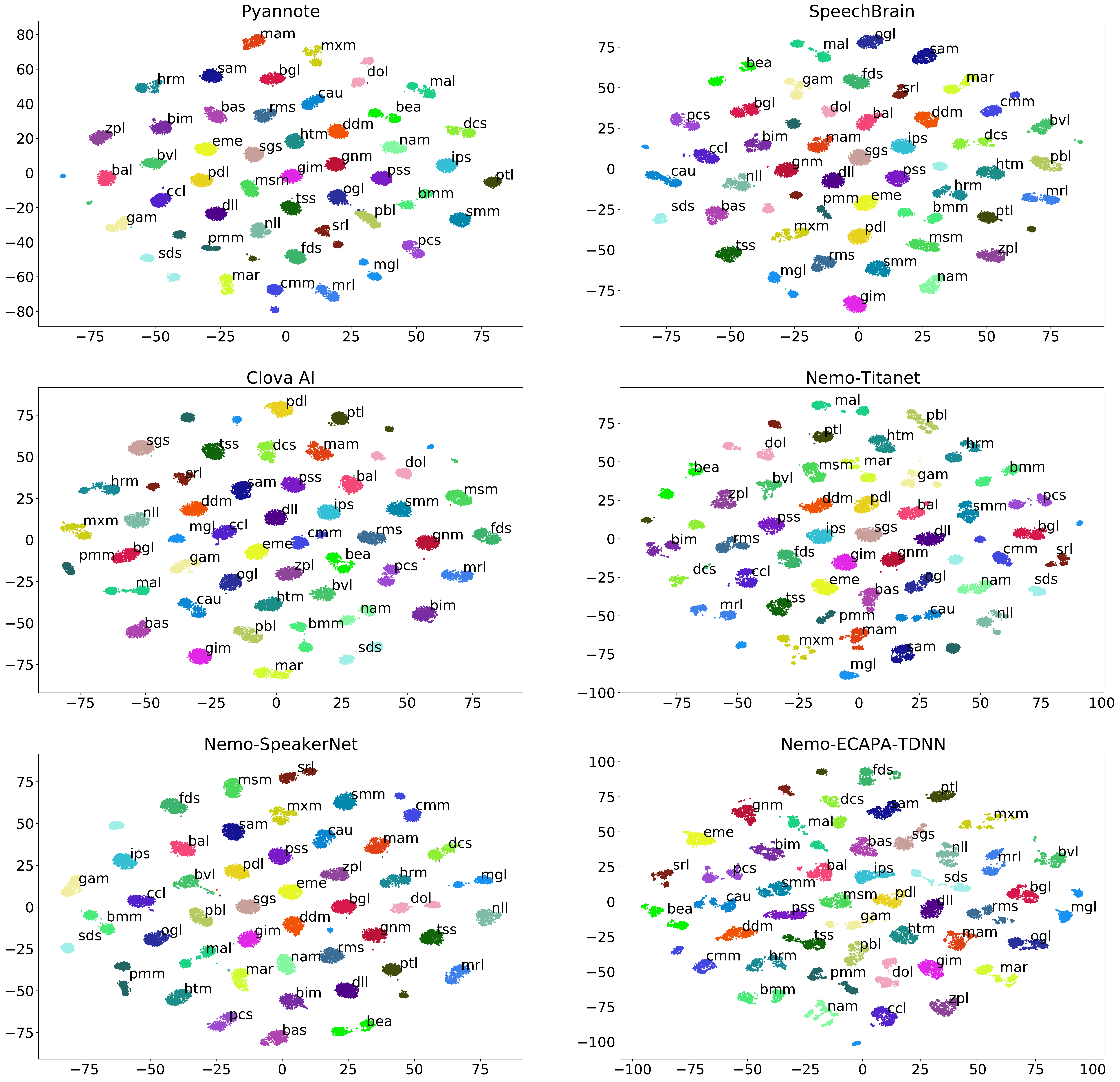}
\caption{t-SNE representations of the speaker embeddings extracted from the different architectures.\label{fig:tsne}}
\end{figure}

Visually examining high-dimensional data is not feasible. However, in most cases, visual representations of the data are more informative to the developer than just percentages and numbers. As such, we employed a t-SNE dimensionality reduction technique~\cite{tsne} and plotted the speaker embeddings obtained from the six architectures into a two-dimensional space. The algorithm was applied over the entire set of embeddings from each individual architecture, ran over 1000 steps and a perplexity of 30 was used. 
Figure~\ref{fig:tsne} shows these t-SNE plots for each architecture. It can be noticed that, with some minor exceptions, the speakers are clustered nicely. Moreover, it does not seem that any of the speaker embedding architectures shows a better performance in terms of grouping the speakers. 
It appears that for some of the speakers, sub-clusters of the embeddings are formed, and in a few cases, outliers are to be observed. By examining the individual speech samples, we noticed that the outliers commonly pertain to the short utterances, reaffirming the previous result of the correlation between the embeddings and the duration of the audio. For the sub-clusters, most speakers recorded the entire set of prompts in at least two recording sessions. Between the sessions, there were differences in the background noise, distance from the microphone, speaking rhythm and vocal effort.
The formed t-SNE clusters are indeed consistent with the different recording sessions. It can be noticed that these clusters are more common within the SWARA2.0 subset containing the home recordings. Among the different embedding architectures, it appears that all of them are able to detect these sub-clusters, while NeMo SpeakerNet and Pyannote seem to be more affected by the duration of the audio and present more outliers than the other systems. This means that these two architectures would be more suitable in pre-processing a  speaker dataset for downstream applications.

\begin{figure}[H]
\includegraphics[width=\columnwidth]{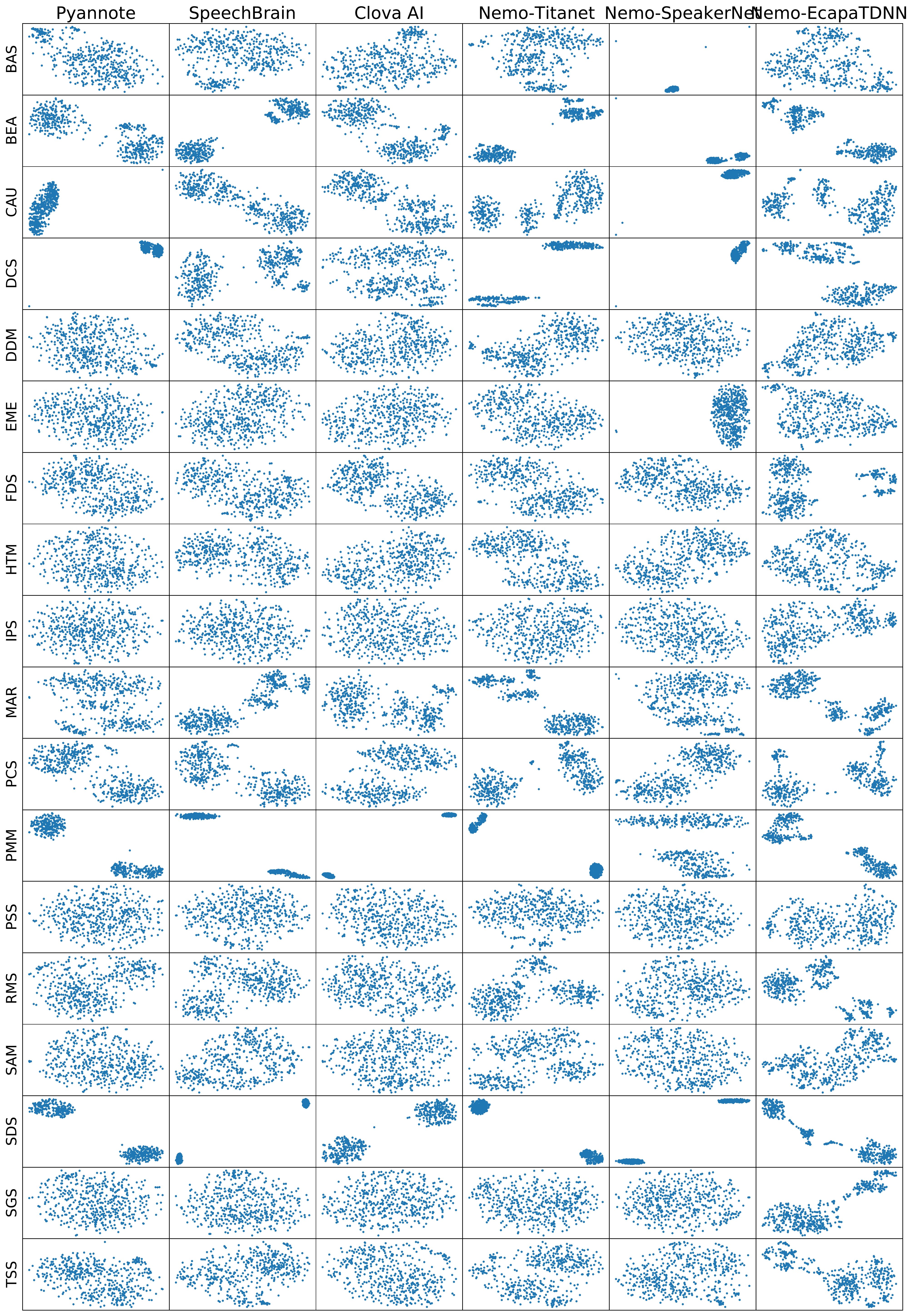}
\caption{Speaker-level t-SNE plots of the different embedding architectures---\textbf{SWARA1.0
} subset.\label{fig:tsne-swara1}}
\end{figure}

\begin{figure}[H]
\includegraphics[width=\columnwidth]{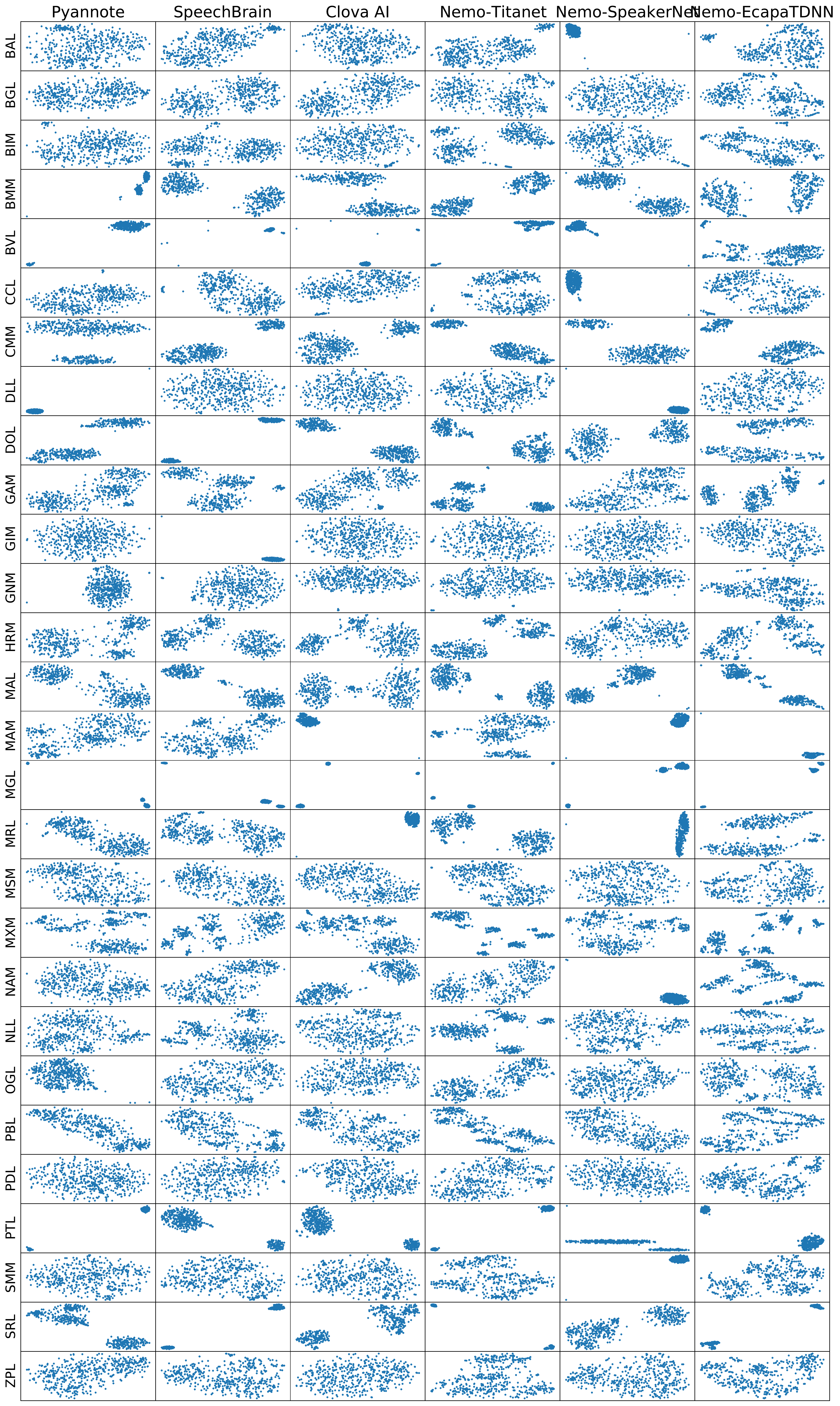}
\caption{Speaker-level t-SNE plots of the different embedding architectures---\textbf{SWARA2.0} subset.\label{fig:tsne-swara2}}
\end{figure}

However, when we zoom in on these low-dimension visualizations, we notice some interesting patterns. Figures~\ref{fig:tsne-swara1} and~\ref{fig:tsne-swara2} show the speaker-level t-SNE representations for the two data subsets, i.e., SWARA1.0 and SWARA2.0, respectively.

\unskip

\clearpage

\clearpage

\section{Conclusions and Future Work}
\label{sec:conc}

In this paper, we attempted to evaluate some of the most recent and high-performing deep speaker embeddings with respect to their intended use, i.e., speaker discrimination, as well as with respect to their drawbacks in terms of residual information present in the embeddings. The selected architectures are Pyannote, Speech Brain, Clova AI, NeMo Titanet, NeMo SpeakerNet and NeMo ECAPA-TDNN, and they were evaluated over a large multi-speaker parallel dataset containing over 38 h of spoken data. 

In a first set of experiments, the architectures were evaluated in terms of the EER and inter- and intra-speaker similarity measures. With respect to the EER, the best discrimination was obtained by NeMo Titanet. However, in terms of the intra-speaker similarity, the architecture which was able to better cluster the speakers was Clova AI, while NeMo ECAPA-TDNN performed best at maximizing the distance between the speakers' representations. This set of evaluations also looked into the different subsets of the audio data, i.e., the male vs. female and studio recordings vs. home recording subsets. The results showed that the female and studio-recorded speakers achieve  lower EER and higher intra-speaker cosine similarity measures. In addition, the male and home-recorded speakers exhibit larger inter-speaker cluster distances. 

A second set of experiments measured the amount of residual information present in the six sets of speaker embeddings. Simple classification and regression algorithms were employed. These algorithms were supposed to achieve high accuracy measures when different speech factors were present in the embeddings. The examined factors were: the utterance duration in terms of the number of characters and signal duration, recording conditions, signal-to-noise ratio and linguistic contents. All six architectures showed high correlations to the length of the signal and recording conditions, including the SNR. The least amount of residual information pertaining to the recording conditions was present in the Pyannote architecture. With respect to the utterance duration and SNR, NeMo Titanet-based embeddings were less correlated to these factors, and NeMo SpeakerNet embeddings had the smallest correlation factor with the linguistic contents of the utterance. However, the differences between the six architectures are minimal, and we posit that, to this point, none of them have truly obtained a disentangled speaker representation.

Given the results of the residual information's presence in the embeddings, a third set of  experiments looked into how these residual factors could be exploited in further downstream applications of the speaker embeddings. Low-dimensional t-SNE-based representations of the six sets of embeddings were plotted. With respect to global speaker representations, all the architectures showed a similar performance with well-behaved clusters, with the exception of NeMo ECAPA-TDNN for which the clusters had larger distributions. When zooming in on these t-SNE representations at the speaker level, all deep representations exhibited sub-clusters pertaining to different recording sessions, as well as outlier utterances correlated to short utterances. This means that this information present in the embeddings' projections could in fact be used, for example, in the data selection process for text-to-speech or voice cloning applications. Outlier utterances could be removed from the training set, and ill-behaved speaker datasets  could be further curated or removed altogether. This is in fact one of the next steps to extend the work presented in this paper. We plan to examine how different text-to-speech architectures are affected by the variability of certain speakers and if removing utterance outliers enhances the performance of the output synthesized speech. Another important result of this work pertains to the use of embedding-based similar speakers for data augmentation in TTS systems, meaning that using the most similar speaker with respect to the target speaker will indeed improve the naturalness and speaker similarity of the resulting system. 

Moreover, given the availability of the speaker embedding networks, we are planning to use the findings of this study in a task of multi-speaker text-to-speech synthesis system training and determine the most efficient manner to input these embeddings into the synthesis networks, as well as to verify how the embeddings are affected by the synthetic output and how they can  be adjusted to better represent the various speaker identities.

\vspace{6pt}



\funding{The work carried out in this paper was partially funded by Zevo Technology under grant agreement no. 21439. The APC was funded by the Technical University of Cluj-Napoca.
}

\institutionalreview{Not applicable}
\informedconsent{Not applicable.}
\dataavailability{The processed datasets and experiment flows used in this paper can be obtained from the author.} 
\conflictsofinterest{The author declares no conflict of interest.} 


\begin{adjustwidth}{-\extralength}{0cm}
\reftitle{References}

\end{adjustwidth}

\end{document}